\documentclass[fleqn,twoside]{article}
\usepackage{espcrc2}
 
\usepackage{graphicx}
\usepackage{epsfig} 
 
\newcommand{\AmS}{{\protect\the\textfont2
  A\kern-.1667em\lower.5ex\hbox{M}\kern-.125emS}}
 
\title{Latest measurements of beauty quark production at HERA
\vspace*{-35mm}{\it
\begin{flushleft} \small
Talk presented at the 31$^\mathit{st}$ International Conference
on High Energy Physics, \\
24-31 July 2002, Amsterdam, The Netherlands.
\end{flushleft}
}\vspace*{21.6mm}
}

\author{V. Chiochia
\address[DESY]{DESY - Deutsches Elektronen-Synchrotron, \\
        Notkestrasse 85, 22607 Hamburg, Germany}}

\begin{document}

\begin{abstract}
The latest results of beauty quark production measurements at HERA are
presented. New measurements have been obtained both in the photoproduction and 
the deep inelastic scattering regimes. The results were compared with the NLO
QCD calculations. 
\vspace{-0.2cm}
\end{abstract}
 
\maketitle

\section{Introduction}
%
%
Production of heavy quarks is a fascinating tool to
understand the nucleon structure as well as the underlying QCD parton dynamics. 
In addition production of $b$ quarks is a source of
background for many searches for new physics at existing and future
colliders. The high $b$ mass ($M_b\gg\Lambda_{QCD}$)
should provide a solid basis for a perturbative calculation.
Nevertheless, so far discrepancies between the experimental cross section
measurements and the NLO QCD calculation have been found.

The experimental procedure often relies on the measurement of the transverse
momentum $p_T^{rel}$ of the muon produced in the semi-leptonic decay 
with respect to the axis of the closest jet. 
This spectrum is harder for $b$ quarks than for $c$ quarks 
and, therefore, allows a statistical separation of the signal from the
background. Additional information such as lifetime measurements provided by
silicon detectors can also be used.
An alternative tag is given by the coincidence of $D^*$ mesons and muons
which provide sensitivity to the region of low transverse $b$ quark momenta and
is less affected by background.

The recent experimental results obtained at the HERA
collider are reviewed in both the photoproduction ($Q^2\sim 0$ GeV$^2$) and 
the Deep Inelastic Scattering (DIS: $Q^2 >1$ GeV$^2$) kinematic regimes.
%
%
\section{Open Beauty in Photoproduction}
\begin{figure}[htb]
\begin{center}
\psfig{file=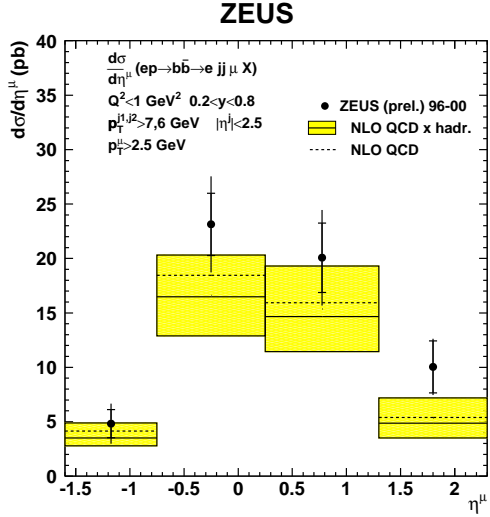,width=7.2cm,clip=,angle=0,silent=}
\end{center} 
\vspace{-1.3cm}
\caption{Differential cross section of beauty photoproduction 
as a function of the muon pseudorapidity compared with NLO QCD calculations.}
\label{FIG:sigmavsetaphp}
\end{figure}

%
%
When the exchanged photon has small virtuality ($Q^2$) the time-scale
of the interaction is such that its hadronic structure can be revealed. 
Photoproduction at HERA can therefore be similar to the 
processes at hadron colliders and supplies complementary information.
Both ZEUS and H1 already published results on $b$ photoproduction
%
\cite{Adloff:1999nr,H1:979,Breitweg:2000nz}.
The ZEUS experiment has now measured the differential cross sections of
beauty photoproduction using events with at least two jets  and a muon in
the final state \cite{ZEUS:785}. The luminosity used is almost three times
larger than in the previous measurements. The fraction of events from $b$
decays has been extracted using the $p_T^{rel}$ method. 
The kinematic region is defined by $Q^2 < 1$ GeV$^2$,
$0.2<y<0.8$, $p_T^{Jet1(2)} > 7(6)$ GeV, $|\eta^{Jet1(2)}|<2.5$,
$p_T^\mu > 2.5$ GeV and $-1.6<\eta^\mu<2.3$
\footnote{$\eta=-\ln(\tan\theta/2)$ is the pseudorapidity, where
$\theta$ is the polar angle measured with respect to the proton beam direction.}.
Figure \ref{FIG:sigmavsetaphp} shows a comparison between the measured
differential cross section and a NLO QCD calculation for different
regions of the muon pseudorapidity. The QCD prediction
was calculated using the program {\tt FMNR} \cite{Frixione:1994dv}. 
The hadronization is modeled by a Peterson function and the spectrum of the semi-leptonic
muon momentum was extracted from {\tt PYTHIA} \cite{Sjostrand:1993yb}. 
The bands around the NLO prediction show the results obtained by varying the $b$ 
quark mass as well as the renormalization and factorization scales.
The measured cross sections are a factor 1.4 larger than the central 
NLO prediction but compatible with it within the experimental and 
theoretical uncertainties.

In addition, a dijet cross section 
($ep \rightarrow b\overline{b}X \rightarrow~jet~jet~X$) 
has been determined using {\tt PYTHIA}
to extrapolate to the unmeasured part of the muon kinematics and to correct 
for the branching ratio. For this measurement a different data sample has
been used with looser cuts on the transverse momentum of
the muon at large pseudorapidities. The result is 
$\sigma^{dijet} = 733 \pm 61 \pm 104$ pb while the NLO QCD prediction
is $381^{+117}_{-78}$ pb, corresponding to a data excess of a factor of
two.
%
%
\section{Open Beauty in DIS}
\begin{figure}[htbp]
\begin{center} 
\psfig{file=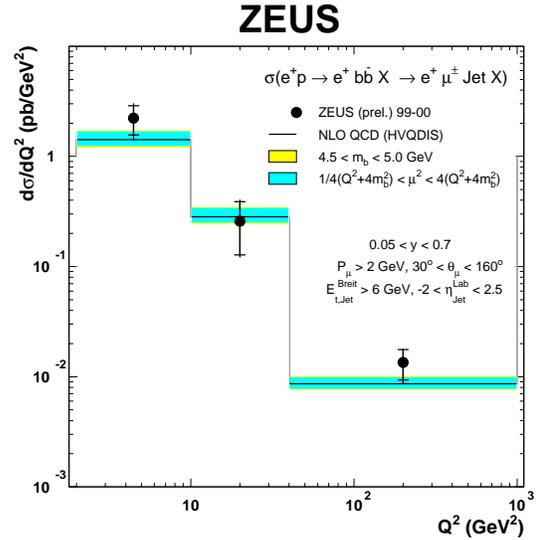,width=6.9cm,clip=,angle=0,silent=}
\end{center} 
\vspace{-1.3cm}
\caption{Differential beauty cross section as a function of $Q^2$ compared
with the NLO QCD calculations.}
\label{FIG:sigmavsq2dis}
\end{figure}

%
%
First results in the DIS region have been already released by the H1 
collaboration \cite{H1:1013}. Thanks to the high luminosity
in the new ZEUS measurement \cite{ZEUS:783} differential distributions
have been measured for the first time in DIS.
Events were selected by
requiring the presence of at least one muon in the final state
and at least one jet in the Breit frame\footnote{In the Breit frame, defined by 
$\vec{\gamma}+2x\vec{P} = \vec{0}$, where $\vec{\gamma}$ is the momentum of the 
exchanged photon, $x$ is th Bjorken scaling variable and $\vec{P}$ is the
proton momentum, a purely space-like photon and a proton
collide head-on.}.
A total visible cross section of 
$\sigma^{vis} = 38.7\pm 7.7^{+6.1}_{-5.0}$ pb
was measured for the reaction 
$ep \rightarrow e b\overline{b}X \rightarrow e~jet~\mu~X$ 
in the kinematic region defined by: 
$Q^2 > 2$ GeV$^2$, $0.05<y<0.7$, $p^\mu > 2$ GeV, $30^\circ <\theta^\mu < 160^\circ$
and one jet in the Breit frame with $E_T^{Breit}> 6$ GeV and $-2 < \eta^{Lab} < 2.5$.
Also for this measurement the $p_T^{rel}$ method was applied.
This result has been compared with a NLO QCD calculation implemented
in the {\tt HVQDIS} program \cite{Harris:1997zq}, after folding the
$b$ quark momentum spectrum with a Peterson fragmentation function and subsequently with a
spectrum of the semi-leptonic muon momentum extracted from {\tt RAPGAP} \cite{Jung:1993gf}.
The NLO QCD prediction is $28.1^{+5.3}_{-3.5}$ pb which agrees 
with the measured value within the errors. The differential
cross section as a function of $Q^2$ compared to the NLO calculation 
is shown in Fig. \ref{FIG:sigmavsq2dis}.
\vspace{-.5cm}
\vspace{-.4cm}
The prediction of the Monte Carlo program {\tt CASCADE} \cite{Jung:2000hk}, 
which implements a calculation based on the CCFM evolution equations \cite{Ciafaloni:1987ur} 
and uses a $k_T$-dependent gluon density, is 35 pb which
is in very good agreement with the measurement.
In addition, the simulation gives a good description of the measured differential
cross sections.
%
%
\section{$D^*-\mu$ correlations}
%
%
The separation of charm and beauty contributions to the signal can
also be performed by exploiting the charge and angle correlations of the $D^*$ 
meson and of the muon in the reaction 
$ep\rightarrow eb\overline{b}X\rightarrow eD^*\mu X$. 
Of particular interest is the configuration
in which the muon and the $D^*$ originate from the same parent
$B$-meson yielding unlike charge sign $D^*-\mu$ pairs
produced in the same hemisphere.

Using this strategy and performing a likelihood fit on the
kinematic distributions H1 has extracted the cross sections
of beauty and charm production in the kinematic region defined by
$p_T^{D^*}>1.5$ GeV, $|\eta^{D^*}|<1.5$, $p_T^\mu > 1$ GeV,
$|\eta^\mu|<1.74$ and $0.05<y<0.75$ \cite{H1:1016}. 
The measured values, which
confirm previous results, are respectively $\sigma^b_{vis}=380\pm 120\pm 130$ pb and
$\sigma^c_{vis}=720\pm 115\pm 245$ pb which are 
well above LO+parton shower Monte Carlo expectations.

A similar analysis has been conducted by ZEUS \cite{ZEUS:784}
whose selection has been optimized for decays of $b$ quarks.
The beauty cross section, measured in a slighty different
phase space ($p_T^{D^*}>1.9$ GeV, $|\eta^{D^*}|<1.5$, $p_T^\mu > 1.4$ GeV,
$-1.75<\eta^\mu<1.3$), is  $\sigma^b_{vis}=214\pm 52^{+96}_{-84}$ pb.
The result is in good agreement with the 
the H1 measurement once the same cuts are applied.
In order to compare the measured cross section with NLO QCD
predictions a photoproduction-enriched sample has been selected
by applying the cuts $Q^2 < 1$ GeV$^2$ and $0.05<y<0.85$. Furthermore,
the measurement is restricted in a $b$ quark rapidity range $\zeta^b <1$
where the distributions of the Monte Carlo program used to extrapolate
agree with the respective {\tt FMNR} spectra within $\pm15 \%$. 
The result for the extrapolated cross section is 
$\sigma_{\gamma p \rightarrow b(\overline{b})X}=15.1\pm 3.9 ^{+3.8}_{-4.7}$ nb
while the NLO prediction of {\tt FMNR} for this reaction is only
$5.0^{+1.7}_{-1.1}$ nb.

\section{Conclusions and outlook}
%
%
The understanding of the $b$ quark production mechanism is 
an outstanding puzzle in QCD. 
A set of new visible cross sections of beauty production, defined 
close to the detector acceptance, have been measured at HERA.
The results are about a factor 1.4 higher than the NLO QCD predictions but
consistent within the experimental and theoretical uncertainties both in
the photoproduction and the DIS regions. An excess of measured cross
sections over the NLO QCD prediction is observed 
when attempts are made to extrapolate cross sections 
to regions which are not directly measured by the detectors.

The HERA collider is now starting a new phase of operation at
higher luminosities. Together with the enhanced $b$-tagging capabilities 
of the new H1 and ZEUS vertex detector and tracking triggers 
more precise and differential measurements can be expected within
the next five years.

\vspace{0.3cm}
{\bf Acknowledgments}\\
I would like to thank my collegues of the H1 and ZEUS collaborations
for their work and their suggestions to this talk.

\end{document}